\documentclass[a4paper,10pt]{article}
\usepackage{amsmath}
\usepackage{a4wide}
\usepackage{psfrag}
\usepackage{boxedminipage}
\usepackage{pstricks,pst-node,pst-text,pst-3d}
\usepackage{amsfonts}
\usepackage{amssymb}
\usepackage{fancyhdr}
\usepackage{graphicx}
\usepackage{float}
\usepackage{epsfig}
\usepackage{geometry}
\usepackage{fancybox}
\usepackage{float}
\usepackage{slashed}
\usepackage{mathenv}
\usepackage{hyperref}
\usepackage{pstricks}

\def\noi{\noindent}

\newcommand{\mchi}{{m_{\chi}}}

\newcommand{\neuto}{{\tilde{\chi}}_1^0}
\newcommand{\mneuto}{m_{{\tilde{\chi}}_1^0}}
\newcommand{\neutt}{{\tilde{\chi}}_2^0}
\newcommand{\mneutt}{m_{{\tilde{\chi}}_2^0}}

\newcommand{\chargop}{{\tilde{\chi}}_1^+}
\newcommand{\chargopm}{{\tilde{\chi}}_1^\pm}
\newcommand{\chargomp}{{\tilde{\chi}}_1^\mp}

\newcommand{\mchargopm}{m_{{\tilde{\chi}}_1^\pm}}
\newcommand{\chargtp}{{\tilde{\chi}}_2^+}
\newcommand{\chargtpm}{{\tilde{\chi}}_2^\pm}
\newcommand{\mchargtp}{m_{{\tilde{\chi}}_2^+}}
\newcommand{\mchargtpm}{m_{{\tilde{\chi}}_2^\pm}}
\def\mchargo{m_{\tilde{\chi}_1^+}}
\def\mchargt{m_{\tilde{\chi}_2^+}}
\def\chargom{\tilde{\chi}_1^-}

\newcommand{\Att}{\mathrm{A}_{\tau \tau}}

\newcommand{\drbar}{{\overline{\rm DR}}}
\newcommand{\mhs}{{MH}}
\newcommand{\atts}{{\mathrm{A}_{\tau \tau}}}
\def\mct{m_{\neuto} m_{\chargop} m_{\chargtp}}

\def\SloopS{\texttt{SloopS}}
\def\micromegas{{\texttt{micrOMEGAs}}}

\def\SuSpect{\texttt{SuSpect}}

\def\tgb{\tan \beta}
\newcommand{\tb}{t_\beta}

\newcommand{\stau}{\tilde{\tau}}

\newcommand{\Omegah}{\Omega_{\chi} h^{2}}

\newcommand{\beqn}{\begin{eqnarray}}
\newcommand{\eeqn}{\end{eqnarray}}
\newcommand{\bea}{\begin{eqnarray}}
\newcommand{\ena}{\end{eqnarray}}
\newcommand{\ra}{\rightarrow}

\begin{document}
\begin{titlepage}
\vspace*{0.1cm}
\rightline{PITHA 09/27}
\rightline{LAPTH-1357/09}

\vspace{1mm}
\begin{center}

{\bf
Relic density at one-loop with gauge boson pair production}

\vspace{.5cm}

{N.~Baro${}^{1)}$, F.~Boudjema${}^{2)}$, G.~Chalons${}^{2)}$, Sun Hao${}^{2)}$ }\\

\vspace{4mm}

{\it 1) Institut f\"ur Theoretische Physik E, RWTH Aachen
University,\\ D-52056 Aachen, Germany} \\
{\it 2) LAPTH, Universit\'e de Savoie, CNRS, \\
BP 110, F-74941 Annecy-le-Vieux Cedex, France}

\vspace{10mm}

\abstract{
We have computed the full one-loop corrections (electroweak as
well as QCD) to processes contributing to the relic density of
dark matter in the MSSM where the LSP is the lightest neutralino.
We cover scenarios where the most important channels are those
with gauge boson pair production. This includes the case of a bino
with some wino admixture, a higgsino and a wino. In this paper we
specialise to the case of light dark matter much below the TeV
scale. The corrections can have a non-negligible impact on the
predictions and should be taken into account in view of the
present and forthcoming increasing precision on the relic density
measurements. Our calculations are made with the help of \SloopS,
an automatic tool for the calculation of one-loop processes in the
MSSM. The renormalisation scheme dependence of the results as
concerns $\tgb$ is studied.

}
\end{center}
\normalsize
\end{titlepage}
\newpage

\section{Introduction}
Existence of nonbaryonic dark matter is established through
precise determination of the mean densities of matter in the
Universe. However one does not know what constitutes dark matter
even if the measurement of the relic density of cold dark matter
is now very precise\cite{WMAP-5yr}. On the other hand the latest
observations dedicated to the detection of dark matter have
recently received a lot of attention: the PAMELA collaboration has
reported a $100$ GeV excess on the ratio of fluxes of cosmic ray
positrons to electrons\cite{Adriani:2008zr} and the ATIC balloon
experiment claims a cut-off in the positron flux near $500$ GeV\cite{ATIC:2008zzr}. The FERMI\cite{FERMI09} and
HESS\cite{Aharonian09} observations do not confirm this data but
still point to a deviation from the power-law spectrum. Many
explanations were advocated to account for these results and the
most optimistic and exciting one is that it could be a signal of
annihilating dark matter. Supersymmetry can provide, among many
other advantages, a dark mater candidate through the lightest
supersymmetric particle (LSP) which is a neutralino, $\neuto$, if
R-parity is conserved. Meanwhile, the search for dark matter will
soon also take place within colliders, in particular the LHC. If
dark matter is discovered among the other new particles that form
a New Physics model, one will be able to probe its properties.
One could then predict the relic abundance of the Universe and
would constrain cosmology with the help of precision
data\cite{baro07,wmaplhclc-requirements,Peskin-DM-requirements}
provided by WMAP\cite{WMAP-5yr} and PLANCK\cite{planck}. The
present WMAP accuracy on the relic density is about 10\% and with
the PLANCK satellite that has been launched recently it will reach
about 2\% precision. Sophisticated codes exist
\cite{micromegas,dark-susy,superiso-relic} for the calculation of
the relic density in supersymmetry with the inclusion of some
higher order effects, essentially through some running
couplings/masses or some effective couplings (particularly
corrections to the Higgs couplings that can drastically change the
results in the so called Higgs
funnel\cite{micromegas,Klasen-relic-qcd} for example). However
these codes are essentially based on tree-level cross sections.
To match the experimental accuracy, on the theoretical side we
have to provide more precise calculations. Therefore we need to
evaluate annihilation and co-annihilation cross sections at least
at next-to-leading order. Considering the very large number of
processes required for the evaluation of the relic density and the
number of diagrams that each process involves, especially at one
loop, automation of the relic density calculations especially in
the MSSM in unavoidable. The purpose of this paper is to present
some results on the one-loop calculation of the relic density of
the LSP in the MSSM, where the dominant annihilation and
co-annihilation channels are dominated by annihilations into gauge
bosons. Beside the physics motivation for such scenarios,
calculations of these processes involving gauge bosons are
challenging. Attempts to include some effects through effective
couplings are tricky because one has to be careful about
maintaining gauge invariance and unitarity. A preliminary study of
such scenarios has been made by some of us\cite{baro07,barosusy09}.
In this paper we consider the case of a relatively light
neutralino. Very heavy neutralinos with TeV masses and above will
be studied in a subsequent paper especially since they
show new interesting effects. \\

\noi The relic abundance will be derived from the assumption that it is
thermally produced in the early stages of the
universe\cite{jungman96}, so in a first approximation the relic
density is inversely proportional to the thermal average cross
sections, $\langle \sigma v \rangle$. The computations that we
present here are performed with the help of
\texttt{SloopS}\cite{boudjema05,baro07,Sloops-higgspaper,baro09},
which is a fully automated code for the NLO calculation of any
cross section or decay in the MSSM. Although our main interest
concerns the channels with gauge bosons in the final state, we
will find that there are non negligible co-annihilations channels
with quarks in the final state. We calculate both the electroweak
and QCD corrections to these contributions. There can be a host of
processes contributing to the relic density for a particular
scenario. In this paper we calculate however the radiative
corrections only to those processes which, at tree-level,
contribute more than $5\%$ to the relic density. We study here
three different scenarios corresponding to three different
compositions of our lightest neutralino: i) a bino like neutralino
with some wino admixture , ii) a higgsino like neutralino, iii) a
light wino like neutralino. We will also study the impact of
different choices of the renormalisation scheme for $\tgb$ on the
corrections. To derive the corrected relic density we feed our
velocity dependent cross sections into {\tt
micrOMEGAs}\cite{micromegas} for performing the thermal average
and solving the Boltzmann equations. We will always show the
cross sections (at tree-level and at one-loop) in terms of the
relative velocity and as a guide we extract the $s$-wave and
$p$-wave coefficients and the corrections they receive. This helps
also extract the, one-loop, Sommerfeld\cite{textbook-Sommerfeld}
factor for some of the cross sections. In the processes we study
here these are of QED origin. Once these one-loop Sommerfeld QED
corrections are extracted we first subtract them before
performing the all order resummation and deriving the relic
density.

\noi The paper is organised as follows. In the next section we give an
overview of {\tt SloopS} and on how we perform the one-loop
calculations, in particular summarising our renormalisation
procedure. Checks on the calculations as concerns ultra-violet
finiteness, infra-red finiteness and gauge parameter independence
are spelled out. The interface between {\tt SloopS} and {\tt
micrOMEGAs} will also be presented. Most of the scenarios that we
will be studying involve co-annihilation, we will define the
effective cross section that includes the statistical weight. At
the end of this section we present how our models have been
defined. Section~3 considers the case of a light (about $100$ GeV)
mixed bino-wino LSP. The case of a dominantly higgsino LSP with
mass of about $200$ GeV is studied in Section~4. Section~5 covers
the case of a light wino of about the same mass. The last section
contains our conclusions and prospects.

\section{Overview of the calculation}
\subsection{Set up of the automatic calculation: \texttt{SloopS}}
One-loop processes calculated via the diagrammatic Feynman
approach involve a huge number of diagrams even for $2 \rightarrow
2$ reactions, especially in a theory like supersymmetry. Doing
full calculations by hand without automation is practically
untractable. There exists already efficient automatic codes for
one-loop calculations\cite{grace-1loop,Grace-susy1loop,formcalc}.
\texttt{SloopS} is an automated code for one-loop calculations in
supersymmetry. It is a combination of
\texttt{LanHEP}\cite{lanhep}, the bundle
\texttt{FeynArts}\cite{feynarts}, \texttt{FormCalc}\cite{formcalc}
and an adapted version of \texttt{LoopTools}\cite{looptools,
boudjema05} (that we will call the \texttt{FFL} bundle from now
on). \texttt{LanHEP} deals with one of the main difficulties that
has to be tackled for the automation of the implementation of the
model file, since this requires that one enters the thousands of
vertices that define the Feynman rules. On the theory side a
proper renormalisation scheme needs to be set up, which then means
extending many of these rules to include counter-terms. This part
is done through \texttt{LanHEP} which allows to shift fields and
parameters and thus generates counterterms most efficiently. The
ghost Lagrangian is derived directly from the BRST
transformations. The loop libraries used in \texttt{SloopS} are
based on \texttt{LoopTools} with the addition of quite a few
routines in particular those for dealing with small Gram
determinants that appear in our case at small relative velocities
of the annihilating dark matter, and even more so of relevance for
indirect detection\cite{boudjema05}.
\subsection{Non-linear gauge fixing}
We use a generalised non-linear gauge\cite{chopin-nlg,boudjema05} adapted
to the minimal supersymmetric model. The gauge fixing writes
\begin{eqnarray}
 {\mathcal L}_{GF} &=& - \frac{1}{\xi_W}
|(\partial_\mu - ie
   {\tilde{\alpha}}  A_\mu - igc_W  {\tilde{\beta}} Z_\mu) W^{\mu \, +}
+ i{{ \xi_W}}  \frac{g}{2}(v +  {\tilde{\delta}} h^0 + {\tilde{\omega}} H^0 + i{\tilde{\kappa}} G^0+ i {\tilde{\rho}}A^0)G^+\mid^2 \nonumber\\ & \quad &
- \frac{1}{2  {{ \xi_Z}}  } (\partial_\mu Z^\mu +
  {{ \xi_Z}}  \frac{g}{2c_W}(v +
   {\tilde{\epsilon}}  h^0 +  {\tilde{\gamma}}_H^0  )G^0)^2  -\frac{1}{2 {{ \xi_A}} } (\partial_\mu A^\mu)^2 \, .
\end{eqnarray}\label{gaugefixing}
Unlike the other parts of the model ${\mathcal L}_{GF}$
is written in terms of {\it renormalised} fields and parameters. $G^0$,$G^\pm$ are
the Goldstone fields. We always work with $\xi_{A,Z,W} = 1$ so as to deal with the minimal
set of loop tensor integrals. This implementation of the gauge fixing is very useful
to check gauge independence for processes with gauge boson production. More details
are given in\cite{Sloops-higgspaper}.
\subsection{Renormalisation}
In our code we have renormalised and implemented each sector of
the MSSM. This is explained in details
in\cite{baro07,Sloops-higgspaper,baro09}. Here we only briefly
sketch the renormalisation procedure. We have worked, as far as
possible, within an on-shell scheme generalising what is done for
the
electroweak standard model\cite{grace-1loop}.\\

\noi {\it {\bf i)}} The Standard Model parameters : the fermion masses
as well as the mass of the W and Z are taken as input physical
parameters. The electric charge is defined in the Thomson limit,
see for example\cite{grace-1loop}. The light quarks (effective)
masses are chosen such as to reproduce the SM value of
$\alpha^{-1}(M_Z^2)$ = 127.77. This should be kept in mind since
one would be tempted to use a $\drbar$ scheme for $\alpha$,
defined as $M_Z$, to take into account the fact that dark matter
is annihilating at roughly the electroweak scale, so that
$\alpha(M_Z^2)$ is a more appropriate choice. We should keep in
mind that doing so would amount to correcting the tree-level cross
section by about $13\%$ for $2 \ra 2$ processes. As we will see
this running does not, most of the time, take into account the
bulk of the radiative corrections that we report here.
\\

 \noi {\it {\bf ii)}} The Higgs sector : We take the pseudoscalar Higgs
 mass $M_A$ as an input parameter and require vanishing tadpoles.
$\tgb$ is defined through several schemes whose impact on the
radiative corrections we will study:
\begin{description}
       \item[~-] a $\drbar$ definition where the $\tgb$ counter-term is defined as a
       pure divergence leaving out all finite parts.
    \item[~-] a process-dependent definition of this counter-term by extracting it
    from the decay $A^0 \rightarrow \tau^+\tau^-$ that we will refer to as $\atts$ for
    short. This definition is a good choice for the gauge independence of the processes.
    \item[~-] an on-shell definition with the help of the mass of the heavy CP Higgs
    $H^0$ taken as input parameter called the MH scheme from now on.
    We have reported elsewhere that this scheme usually
    introduces large radiative corrections.

\end{description}
These schemes are thoroughly discussed in
\cite{Sloops-higgspaper}, in particular the question of gauge
invariance of these schemes is addressed.\\

\noi {\it {\bf iii)}} The sfermion sector : For the slepton sector we use as
input parameters masses of the two charged sleptons which in the case of no-mixing
define the R-slepton soft breaking mass, $M_{{\tilde e}_{R}}$ and the $SU(2)$ mass,
 $M_{{\tilde e}_{L}}$, giving a correction to the sneutrino mass at one-loop.
In the squark sector each generation needs three physical masses to constrain
the breaking parameter $M_{{\tilde Q}_{L}}$ for the $SU(2)$ part, $M_{{\tilde u}_{R}}$,
$M_{{\tilde d}_{R}}$ for the R-part. See\cite{baro09} for details.\\

\noi {\it {\bf iv)}} The chargino/neutralino sector :
For this sector we implement an on-shell scheme by taking as input
three masses in order to reconstruct the underlying parameters
$M_1,M_2,\mu$.
In \SloopS~\cite{baro09} the default scheme is to choose two
charginos masses $\mchargopm$ and $\mchargtpm$ as input to define
$M_2$ and $\mu$ and one neutralino mass, $\mneuto$, to fix
$M_1$. The masses of the remaining three neutralinos receive
one-loop quantum corrections. In this scheme, these counterterms
are~\cite{baro09}
\begin{eqnarray}
\delta M_{2}&=&\frac{1}{M_{2}^{2}-\mu^{2}}\left(
(M_{2}\mchargo^2-\mu {\rm det} X)\frac{\delta\mchargo}{\mchargo}
+ (M_{2}\mchargt^2-\mu {\rm det} X)\frac{\delta\mchargt}{\mchargt}\right.\nonumber\\
& &-\left. M_{W}^2(M_{2}+\mu s_{2\beta})\frac{\delta
M_{W}^2}{M_{W}^2} - \mu M_{W}^{2} s_{2 \beta}c_{2 \beta}
 \frac{\delta t_\beta}
{t_{\beta}}\right),\nonumber\\
\delta \mu &=&\frac{1}{\mu^{2}-M_{2}^{2}}\left(
(\mu\mchargo^2-M_{2} {\rm det} X)\frac{\delta\mchargo}{\mchargo}
+ (\mu\mchargt^2- M_{2}{\rm det} X)\frac{\delta\mchargt}{\mchargt}\right.\nonumber\\
& &-\left. M_{W}^2(\mu+M_{2} s_{2\beta})\frac{\delta
M_{W}^2}{M_{W}^2} - M_{2} M_{W}^{2} s_{2 \beta} c_{2 \beta}
\frac{\delta t_\beta}
{t_{\beta}}\right),\\
\delta M_{1}&=&\frac{1}{N_{11}^{*\,2}}(\delta
m_{\chi_{1}^{0}}-N_{12}^{*\, 2}\delta
M_{2}+2N^{*}_{13}N^{*}_{14}\delta\mu\nonumber
\\& &-2N^{*}_{11}N^{*}_{13}\delta
Y_{13} - 2N^{*}_{12}N^{*}_{13}\delta Y_{23} -
2N^{*}_{11}N^{*}_{14}\delta Y_{14} - 2N^{*}_{12}N^{*}_{14}\delta
Y_{24} )\, , \label{dm1dm2dmu}
\end{eqnarray}
with $\textrm{det}X=M_{2}\mu-M_{W}^{2}s_{2\beta}$, $Y$ is the
neutralino mixing matrix and $N$ its diagonalising unitary matrix,
see~\cite{baro09}. Looking at these equations some remarks can be
made. First, in the special configuration $M_2 \sim \pm\mu$ an
apparent singularity might arise. Ref.~\cite{baro07} pinpointed
this configuration which can induce a large $\tb$-scheme
dependence in the counterterms $\delta M_{1,2}$ and $\delta\mu$
and therefore to the annihilation of the LSP into $W$'s for a
mixed LSP, see also~\cite{baro09}. Second, the choice of $\mneuto$
as an input parameter is appropriate only if the lightest
neutralino is mostly bino or if the bino like neutralino is not
too heavy compared to other neutralinos. It is however very easy
to switch to another scheme or choice of input parameters in the
chargino/neutralino sector. For instance if the bino like
neutralino is the NLSP with mass $\mneutt$, like what occurs in
the wino scenario that we study in this paper, we simply take
$\mneutt$ as input in which case $\delta M_{1}$ is obtained from
Eq.~(\ref{dm1dm2dmu}) by $\delta m_{\chi_{1}^{0}} \ra \delta
m_{\chi_{2}^{0}}$ and $N_{1j}
\ra N_{2j}$.\\

\noi {\it {\bf v)}} Finally diagonal field renormalisation is fixed by demanding the
residue at the pole of the propagator of all physical particles to
be unity, and the non-diagonal part by demanding no-mixing between
the different particles when on shell. This is implemented in all
the sectors. \\

\noi {\it {\bf vi)}} Dimensional reduction is used as implemented in the
\texttt{FFL} bundle at one-loop through the equivalent constrained
dimensional renormalisation\cite{CDR}.

\subsection{Infrared divergences}
For the processes $\chi\chi \rightarrow X Y $, $X,Y=W^\pm, Z^0,
f,..$, we can decompose the one-loop amplitudes in a virtual part
$\mathcal{M}_{1loop}^{EW}$ (for co-annihilation processes with
external quarks we also need to add $\mathcal{M}_{1loop}^{QCD}$)
and a counter-term contribution $\mathcal{M}_{CT}$. The sum of
these two amplitudes must be ultraviolet finite and gauge
independent. Due to the virtual exchange of the massless photon
and gluon, this sum can contain infrared divergencies. This is
cured by adding a small mass to the photon and/or gluon,
$\lambda_{\gamma}$ and $\lambda_g$. This is a valid
regularisation, even for QCD, for all the processes we study here
where the non-Abelian nature of QCD does not show up. This mass
regulator should exactly cancel against the one present in the
final state radiation of a photon(gluon). The QED(QCD)
contribution is therefore split into two parts : a soft one where
the photon(gluon) energy $E_{\gamma,g}$ is integrated to less than
some small cut-off $k_c$ and a hard part with $E_{\gamma,g} >
k_c$. The former requires a photon(gluon) mass regulator. Finally
the sum
$\mathcal{M}_{1loop}^{EW+QCD}+\mathcal{M}_{CT}+\mathcal{M}_{\gamma,g}^{soft}(E_{\gamma,g}
< k_c)+ \mathcal{M}_{\gamma,g}^{hard}(E_{\gamma,g} > k_c)$ should
be ultraviolet finite, gauge invariant, not depend on the mass
regulator and on the cut $k_c$. We take the strong coupling
constant at the electroweak scale
$\alpha_s$=$\alpha_s(M_Z^2)$=0.118.

\subsection{Checking the result}
\noi {\it {\bf i)}} For each process and set of parameters, we
first check the ultraviolet finiteness of the results. This test
applies to the whole set of virtual one-loop diagrams. The
ultraviolet finiteness test is performed by varying the
ultraviolet parameter $C_{UV}=1/\varepsilon$, $\varepsilon$ is the
usual regulator in dimensional reduction. We vary $C_{UV}$ by
seven orders of magnitude with no change in the result. We content
ourselves with double precision.\\

\noi {\it {\bf ii)}} The test on the infrared finiteness is
performed by including both the loop and the soft bremsstrahlung
contributions and checking that there is no dependence on the
fictitious photon mass $\lambda_{\gamma}$ or gluon mass
$\lambda_g$.\\

\noi {\it {\bf iii)}} Gauge parameter independence of the results
is essential. It is performed through the set of the {\em eight}
gauge fixing parameters defined in Eq.~(\ref{gaugefixing}). The use
of the eight parameters is not redundant as often these parameters
check complementary sets of diagrams. It is important to note
that in order to successfully achieve this test one should not
include any width in the propagators. However we encountered a $W$
boson resonance for the calculation of $\chi\chi \rightarrow qq'$
and we had to include a width to the W propagator to avoid
numerical instabilities; nevertheless this has been done
\textit{only} for the evaluation of the hard emission part and
\textit{not} for the virtual and soft part. This will be discussed
at more length in due course.\\

\noi {\it {\bf iv)}} For the bremsstrahlung part, the soft
component is added to the virtual corrections and, for the hard
one, we use VEGAS\cite{vegas} adaptive Monte Carlo integration
package provided in the {\tt FFL} bundle and verify the result of
the cross section against {\tt CompHep}\cite{comphep}. The hard
part is also the trickiest, especially when threshold or
resonances are encountered as stated above, so for some
calculations we use BASES\cite{bases} provided in the {\tt GRACE}
package\cite{grace} which have a better treatment of
singularities\cite{grace-1loop}. We choose $k_c$ small enough and
check the stability and independence of the result with respect to
$k_c$.

\subsection{Effective weighted cross sections}
All cross sections $\sigma_{ij}$ where $i,j$ label the
annihilating and co-annihilating DM particles $i,j$ can, in
general, be expanded in terms of the relative velocity $v_{ij}$,
which for neutralino annihilation is
$v=2\beta=2\sqrt{1-4\mneuto^2/s}$. Away from poles and thresholds,
it is a good approximation to write $\sigma_{ij}v_{ij}=a_{ij}+
b_{ij} v^2$, keeping only the $s$-wave, $a_{ij}$, and $p$-wave,
$b_{ij}$ coefficients. With $T$ being the temperature,
$x=\mneuto/T$, the thermal average gives
\beqn
\label{thermalav}
 \langle \sigma_{ij} \; v_{ij} \rangle=a_{ij}+6
(b_{ij}-a_{ij}/4)/x.
\eeqn
With $g_0=2$ the neutralino LSP spin degree of freedom (sdof), the
co-annihilating particle of sdof $g_i$ and mass $m_i$ contributes
an effective relative weight of
\beqn
\label{gieff}
g_{i,eff}=\frac{g_i}{g_0} (1+\Delta m_i)^{3/2} \exp
(-x \; \Delta m_i), \quad \Delta m_i=(m_i-\mneuto)/\mneuto.
\eeqn
\noi The total number of sdof is $g_{eff}=\sum_i
g_{i,eff}$. An approximation to the relic density is obtained
through a simple one dimensional integration
\beqn
\label{relic-app-coann} \Omega h^{2} &=&
\left(\frac{10}{\sqrt{g_*(x_F)}} \; \frac{x_F}{24} \;  \right)
\frac{0.237 \times 10^{-26} {\rm cm}^{3}.{\rm s}^{-1}}{x_F \; J},
\quad
J= \int_{x_F}^{\infty } \langle \sigma v \rangle_{eff} dx/x^2 \nonumber \\
\langle \sigma v \rangle_{eff}&=&\sum_{ij} \frac{g_{i,eff}
g_{j,eff}}{g_{eff}^2}\; \langle \sigma_{ij} \; v_{ij} \rangle.
\eeqn
$a_{ij}, b_{ij}$ that are needed to compute $\sigma_{ij}$ in
Eq.~(\ref{relic-app-coann}) are given in ${\rm cm}^{3} {\rm
s}^{-1}$. $x_F$ represents the freeze-out temperature. $g_*(x_F)$
is the effective degrees of freedom at freeze-out. $g_*$ is
tabulated in {\tt micrOMEGAs} and we read it, as well as $x_F$,
from there. The weight of a channel (see the percentages we will
refer to later) corresponds to its relative contribution to $J$.
We find it instructive to consider the weighted cross section
\beqn
\label{normgef} \frac{g_{i,eff} g_{j,eff}}{g_{eff}^2}\;
\sigma_{ij} \; v_{ij}
\eeqn
By doing this we somehow normalise the contributions of, in
particular, the co-annihilation cross sections which can be very
large compared to the annihilation cross sections, but which at
the end do not contribute as much because of the Boltzmann factor,
$\exp (-x \; \Delta m_i)$. In our plots the weight and statistics
factors are chosen at freeze-out with $x=x_{F}$, see
Eq.~(\ref{gieff}), and for ease of notation we drop the label
$_{eff}$. Since $x=x_{F}$ is the lowest value of $x$, see
Eq.~(\ref{relic-app-coann}), that enters the calculation of the
relic density, the weight factor tends to enhance the real
contribution of the co-annihilation channels. The correct overall
weight is in our case given by \micromegas. This fact should be
taken into account when we compare the figures where the weighted
cross sections are shown and the tables where
the overall weight (extracted from \micromegas) is given.\\
Let us stress once more that in order to derive the relic density
we do not rely on the approximations given in
Eqs~(\ref{thermalav},~\ref{gieff},~\ref{relic-app-coann}) but pass
all the cross sections to \micromegas.

\subsection{Interfacing \SloopS~with \micromegas}
In order to evaluate the relic density, we interfaced \SloopS~with
\micromegas~to take full advantage of its powerful features
concerning the cosmology related part (solving the Boltzmann
equations with co-annihilation, calculation of the effective
degrees of freedom, thermal averaging,..). The connection between
the two codes is summarized in the following:
\begin{enumerate}
\item The MSSM default directory of \micromegas~uses
\SuSpect~\cite{djouadi02}. In so doing it inherits some of the
radiative corrections in particular in the spectrum (mass)
calculation used in \SuSpect~. From the corrected spectrum
\micromegas~works out new effective tree-level underlying
parameters so that gauge invariance is maintained. For the
interface we have removed this default option of reading from
\SuSpect~and created a model file based on the same tree-level
lagrangian as the one used in \SloopS. In so doing both \SloopS~
and \micromegas~calculate the same tree-level cross sections.
This is also a check on our tree-level cross sections.
\item The one-loop cross sections of \SloopS, appearing into the form of
tables showing the cross section as a function of the relative
momentum $p$, $\sigma (p)$, are then interpolated and passed to
\micromegas~which substitutes these new corrected cross sections
to the corresponding tree-level cross sections. Processes that are
not corrected (and hence are not substituted) are of course kept
in the list of processes for the evaluation of the relic density.
\end{enumerate}

\subsection{Finding scenarios in the MSSM parameter space}
The latest limits on the relic density coming from WMAP five years
data give the 2$\sigma$ range\cite{WMAP-5yr},
\begin{equation}
 0.098 < \Omega_{\chi} h^2 < 0.122.
\end{equation}
When $\mchi > m_W$, channels with gauge bosons in the final state
open and LSP's are annihilating very efficiently, making it
difficult to obtain an absolute value of the relic density within
the WMAP bounds. In the MSSM this is realized with a neutralino
which is mostly wino like or higgsino like and its corresponding
mass must be around 1 TeV for the latter and 2 TeV in the wino
case to be in the cosmologically interesting region. Keeping this
in mind we did not worry too much about the value of the relic
cosmic abundance and, instead, we restricted ourselves to get
gauge bosons in the final state to study the origin of large
corrections, if any. Moreover, since the impact of radiative
corrections can be large, there is not so much sense in picking up
a model based on its agreement with the current data on the basis
of a tree-level calculation and finally we could argue that this
agreement can be obtained with non-thermal dark matter production,
or any other mechanism which could avoid too much depletion. This
said we also wanted to have a rather light spectrum as a
supersymmetric solution to the hierarchy problem requires a
relatively light LSP and in order to have scenarios testable at
colliders. Regardless of these remarks we used \micromegas~as a
guide, being careful about translations of effective couplings and
input parameters.
\\
\noindent Last but not least, it is important to stress that we
did not apply radiative corrections to all subprocesses but only
to the ones contributing more than 5\% to the relic density, the
remaining ones were included only at tree-level. Most often the
processes that we do not correct add up to more than $25\%$ of all
the processes contributing to the relic density, even if
individually their weight is small. Therefore when calculating the
correction to the relic density, the one-loop corrections we
compute can get diluted especially if some cancelations occur at
one-loop between the processes we consider. This point should be
kept in mind when we quote the one-loop corrected relic density.
Ideally we should have corrected all cross sections. This could of
course be done with our code {\tt SloopS} and interface to
\micromegas~, however in these exploratory investigations our aim
is to uncover the salient features of the radiative corrections to
annihilation and
co-annihilation of dark matter in supersymmetry.\\

\noi For all the scenarios we will give below, the low energy
tree-level input parameters are defined at the electroweak scale
and are: $M_1$ the $U$(1) gaugino mass, $M_2$ the $SU$(2)
counterpart, $\mu$ the Higgsino ``mass'', $M_3$ the gluino mass,
$M_A$ the mass of the pseudoscalar Higgs boson and $\tgb$. When
not specified we will take a common sfermion mass. The sfermion
trilinear parameter $A_f$ is set to zero for all generations. We
do not impose any gaugino mass unification at the GUT scale. Now
let us describe the scenarios we study:
\begin{description}
 \item [~i)] Mixed-bino scenario: Usually, assuming gaugino mass
 universality at the GUT scale leads to a bino like LSP. This
 gives a relic density which overcloses the universe.
Relaxing this assumption by adding a substantial wino component
one
can match the WMAP range thanks to the opening of gauge boson channels
and co-annihilation with $\chargopm$. This is easily achieved with $M_1 \sim M_2$.
In our fist scenario $M_1 \sim 100$ GeV so that
the mass of the LSP is around $100$ GeV.
\item [~ii)] Higgsino scenario : A mainly higgsino neutralino of mass $\mchi >
m_W$ (here $\mneuto \sim 230$ GeV) will automatically annihilate
dominantly into gauge bosons and, because of the degeneracy with
the lightest higgsino like chargino, co-annihilation takes place
also.
\item [~iii)] Light-wino scenario : A simple way to obtain gauge bosons in
the final states of annihilating neutralinos is to increase its
$SU$(2) type coupling by decreasing the value of $M_2$ in order to
have a wino like neutralino whose mass $\mneuto$ is taken around
$200$ GeV in this case. Once again we have co-annihilation of the
$\neuto$ with the $\chargopm$ because of the small mass gap
between them.
\end{description}

\section{A mixed bino scenario}
The first example we examine corresponds to a neutralino LSP which
has a substantial bino component. It is known that an almost pure
bino does not annihilate enough to give the right relic density in
the radiation dominated standard scenario. As it couples mostly to
particles with largest hypercharge, the R-sleptons, one can
increase the LSP annihilation rate by lowering the R-sleptons
mass. This is typical of the so-called bulk region of mSUGRA. One
can also rely on co-annihilation with the next-to-lightest
supersymmetric particle (NLSP) to reduce the relic abundance. An
example is co-annihilation with the $\stau$. We have studied these
scenarios in \cite{baro07} including one-loop effects. Another
solution is to add some wino component by fixing $M_2$ close to
$M_1$, hence the LSP/NLSP will have strong couplings with the $W$
boson which will dominate the annihilation rate. This is the case
we study here. We take chargino masses within the LEP limits while
all other particles (sfermions, other neutralinos) are heavy. The
underlying parameters of the models are given in
Table~\ref{tab:SUSY1}.
\begin{table}[h!]
\begin{center}
\begin{tabular}{ccccccccc}
 \hline
Parameter & $M_1$ & $M_2$ & $\mu$ & $\tb$ & $M_3$ & $M_{\tilde{L},\tilde{Q}}$ & $A_i$ & $M_{A^0}$ \\
 \hline
Value     & 110   &134.5  & -245  & 10    & 600   & 600                       & 0     & 600 \\
\hline
 \end{tabular}\caption{\em Mixed-bino scenario: Values of the first
 SUSY set of input parameters. Masses are in GeV.}\label{tab:SUSY1}
\end{center}
\end{table}

\noi $M_{1,2}, \mu$ are reconstructed from $m_{\neuto}$ and
$m_{\chargopm}, m_{\chargtpm}$. The relevant physical masses are
$m_{\neuto} = 106.9$ GeV, $m_{\chargopm} = 124.6$ GeV and
$\mchargtp = 274.8$ GeV. At tree-level $m_{\neutt} = 125.3$ GeV.
The neutralino composition is: $\tilde{\chi}_{1}^{0}$ =
$0.94\tilde{B} - 0.20\tilde{W} - 0.27\tilde{H}_{1}^{0} -
0.10\tilde{H}_{2}^{0}$, where $\tilde{B},
\tilde{W},\tilde{H}_{1,2}^{0}$ denote the $U(1)$ gaugino or bino,
the $SU(2)$ gaugino of wino and the higgsino respectively. The
wino component is not negligible. As a consequence annihilation
into gauge bosons is dominant. The main process
$\neuto\neuto\rightarrow W^{+} W^{-}$ contributes $44\%$ to the
relic density. The important co-annihilation channels are
$\neuto\chargop \rightarrow Z^{0} W^{+}$, $\neuto\neutt
\rightarrow W^{+} W^{-}$ both contributing $5\%$ and
$\neuto\chargop \rightarrow u \bar{d}$ contributing $8\%$.
$\neuto\chargop \rightarrow c \bar{s}$ contributes as much as the
$u \bar{d}$ final state. In the following we will refer to only
one of these quark final states, of course both are counted for
the calculation of the relic density. For the the $\neuto\chargop$
co-annihilation, the s-channel exchange of a $W^+$ boson is
dominant, see also \cite{Baer:2005zc}.

\begin{figure*}[htbp]
\begin{center}
\includegraphics[width=0.40\textwidth]{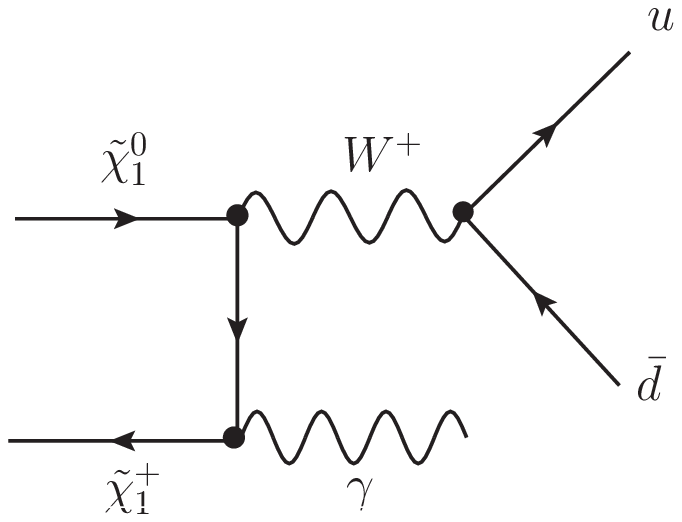}
\includegraphics[width=0.40\textwidth]{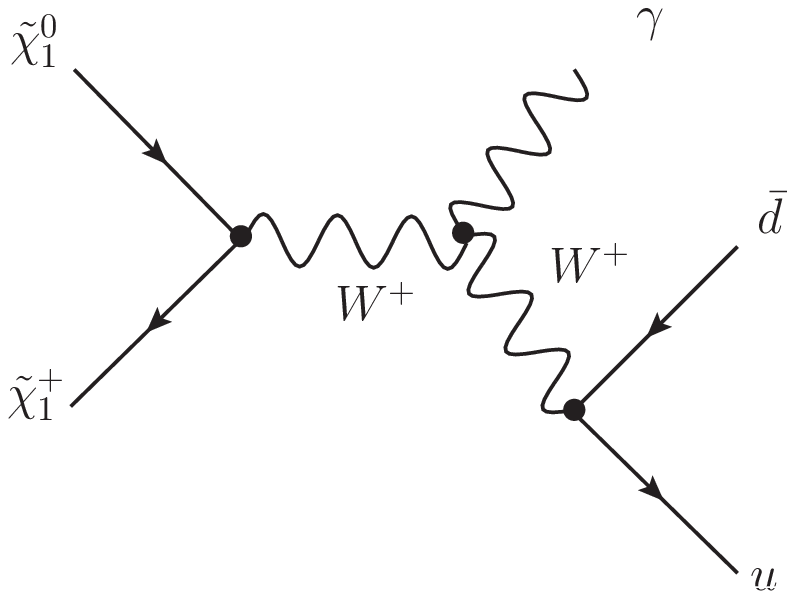}
\caption{\label{figs.diags} \em Real photon emission leading to
$W$ return.}
\end{center}
\end{figure*}

\noi Before showing our results let us comment on a technicality
related to the contribution of the hard bremsstrahlung
contribution. This concerns the radiative process $\neuto \chargop
\rightarrow u \bar d \gamma$, see Fig.~\ref{figs.diags}.

\noi As explained above, when charged/colored particles are involved in
the initial/final state, initial/final state radiation should be
incorporated to have an infrared safe cross section. This emission
is split into two pieces, soft and hard, and the cross section
must be independent of the cut where these two parts are defined.
Calculating the real emission is a tricky task, especially here.
Indeed hard photon emission leads to $W$-return, bringing the
intermediate $W$ on-shell, and therefore would lead to numerical
instability if no width, $\Gamma_W$, is provided for the internal
$W$. We have dealt with this problem by providing a width to the
$W$ only when the radiation is hard, as needed. For the soft part
no width is introduced in order to achieve the cancelation of the
infrared divergence between the soft bremsstrahlung and the
virtual correction where all masses are real. In summary since the
$W$ resonance turns up at an energy much larger than the cutoff
energy $k_c$, where the matching between the soft and hard
emission is done, we decided to split  the integration on the hard
photon phase-space in the $2\rightarrow 3$ process as follows:
\begin{description}
 \item [~i)] from $k_c < E_{\gamma,g} < \frac{1}{2\sqrt s}( s - (M_W^2 + 2\Gamma_W
 M_W))$,
no width is implemented
 \item [~ii)]  from $\frac{1}{2\sqrt s}( s - (M_W^2 + 2\Gamma_W M_W)) < E_{\gamma,g}
 < \frac{1}{2\sqrt s}( s - M_{u \bar d}^2)$ with a width to the W propagator.
\end{description}
One must note that for ii) the hard emission is in fact already
included in the tree-level process $\neuto\chargop \rightarrow W^+
\gamma$ with the $W$ decaying  into a $u \bar d (, c\bar s)$ pair.
The  process $\neuto\chargop \rightarrow W^+ \gamma$ contributes
to the relic density but we did not added to our list of cross
sections to correct as it contributes less than $5\%$. To avoid
double counting when calculating the relic density at one-loop, we
therefore subtract from the list of uncorrected tree-level
contributions $\neuto\chargop \rightarrow W^+ \gamma$ with the
proper branching fraction into $u \bar d, c \bar s$. We will
encounter this feature for all other scenarios that lead to such a
final state and we will treat it in the same way.

\begin{figure*}[htbp]
\begin{center}
\includegraphics[width=\textwidth]{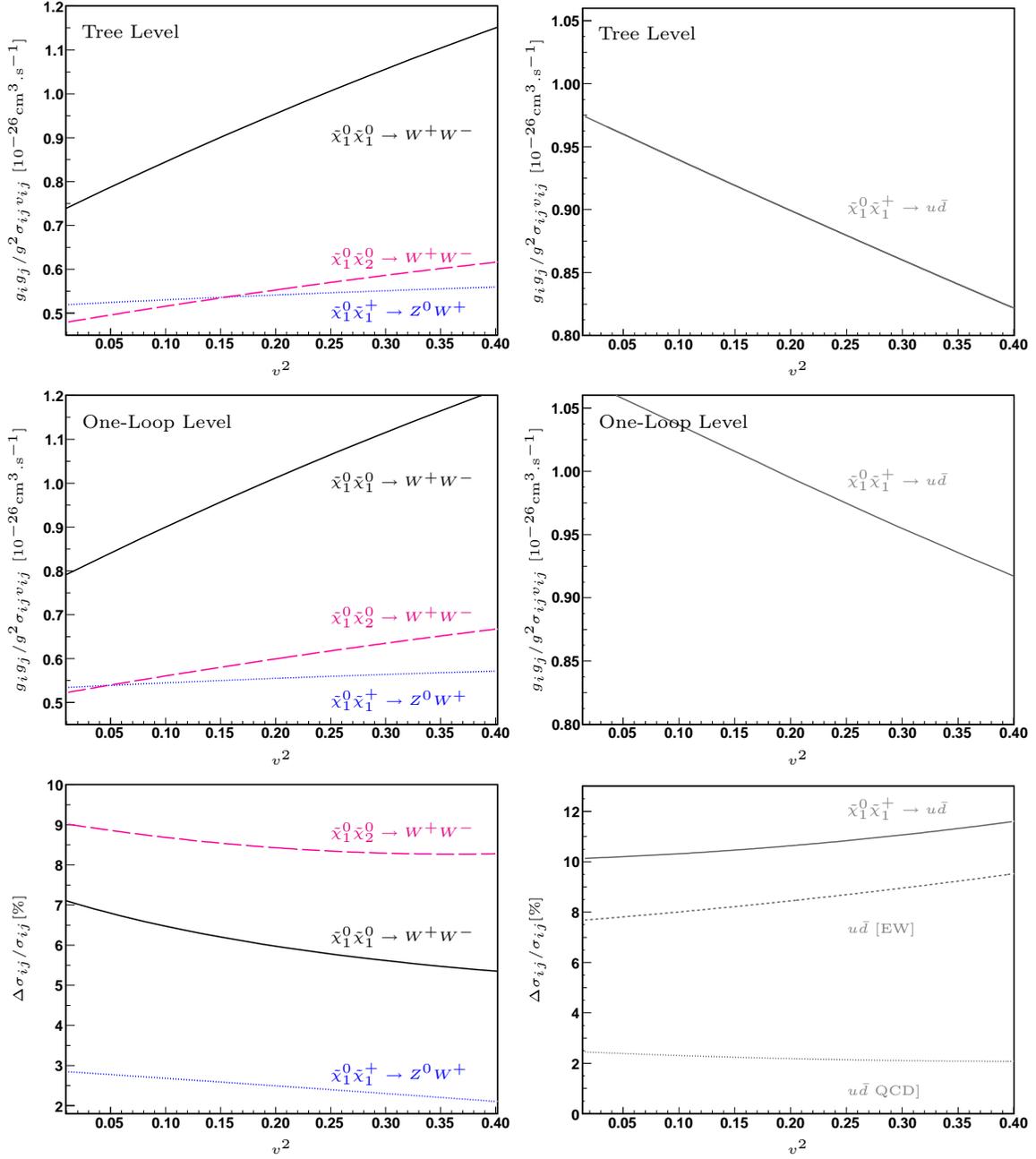}
\caption{\label{figsmixbinogauge} \em Mixed-bino scenario. The
left/right panel shows the main gauge boson/quark production
cross sections respectively. All the cross sections are normalised
with the corresponding effective degrees of freedom given by
Eq.~(\ref{normgef}) with $x_{F}=25.0$. Results are shown for the
$A_{\tau \tau}$ scheme of $\tgb$. }
\end{center}
\end{figure*}

\noi Another point is related to processes which are initiated through
$\neutt$ co-annihilation, $\neuto\neutt \rightarrow W^{+} W^{-}$.
We are working with a scheme where the input parameters are
$\mneuto,\mchargopm,\mchargtpm$. Therefore $\mneutt$ receives a
correction at one-loop. In principle the full one-loop amplitude would write
\begin{eqnarray}
 \mathcal{M}_{\rm 1-loop}(m^{\textrm{one-loop}}_{\tilde{\chi}_{2}^{0}})
= \mathcal{M}_{\rm 1-loop}(m_{\tilde{\chi}_{2}^{0}}) + \Delta
m_{\tilde{\chi}_{2}^{0}} \frac{\partial \mathcal{M}_{\rm
tree}}{\partial m_{\tilde{\chi}_{2}^{0}}}
(m_{\tilde{\chi}_{2}^{0}}) \, ,
\end{eqnarray}
where $\mathcal{M}_{\rm tree}(m_{\tilde{\chi}_{2}^{0}})$ is the
tree-level amplitude, $m_{\tilde{\chi}_{2}^{0}}$ and
$m^{\textrm{one-loop}}_{\tilde{\chi}_{2}^{0}} =
m_{\tilde{\chi}_{2}^{0}} + \delta m_{\tilde{\chi}_{2}^{0}}$ is the
corrected mass. We have neglected the second contribution. This is
because the correction to $m_{\tilde{\chi}_{2}^{0}}$ is less than
$0.3\%$ for all $\tb$ schemes as shown in
Table~\ref{corr-mass-bino}.
\begin{table}[h!]
\begin{center}
\begin{tabular}{rlccc}
\hline
\multicolumn{2}{c}{Masses [GeV]} & $m_{\neutt}$ & $m_{\chi_3^0}$ & $m_{\chi_4^0}$    \\
\hline
\multicolumn{2}{c}{Tree Level} & 125.3 & 258.1 & 270.4  \\
\hline
One Loop & - $\atts$ scheme & 125.13  & 258.58    & 270.42 \\
     & - $\mhs$ scheme & 125.31 & 258.05 & 270.65  \\
         & - $\drbar$ scheme & 125.17  & 258.46 & 270.47  \\
\hline
\end{tabular}\caption{\label{corr-mass-bino}
{\em Mixed-bino scenario: One-loop corrections to the
chargino/neutralino masses in GeV in the scheme $\mct$
for different $t_{\beta}$-schemes: $\atts$, $\drbar$ and $\mhs$.}}
\end{center}
\end{table}
When calculating the relic density we should also in principle use
the corrected physical mass, like for example in the Boltzmann
factor, however again this is negligible. Results for the weighted cross sections at tree-level and at
one-loop are displayed in Fig.~\ref{figsmixbinogauge}.
First of all the QCD and EW corrections to the co-annihilation
into light quarks add up to about $10\%$ and are practically
velocity independent, especially the QCD corrections. The full
${\cal {O}} (\alpha)$ correction to gauge boson production shows
the same feature. The dominant annihilation channel into $W^+W^-$
gets about $+7\%$ correction, while $\neuto\neutt \rightarrow
W^{+} W^{-}$ is slightly larger with $9\%$. $\neuto\chargop
\rightarrow Z^{0} W^{+}$ is small with about $2\%$. Had we used a
running $\alpha$ at $M_Z$ some of the largest positive corrections
would have been absorbed, however our results show that the full
corrections are necessary in view of the upcoming precision on the
extraction of the relic density. For this scenario the
corrections, in the $\alpha(0)$ schemes are positive for all
processes we have considered. Nonetheless, since the wino
component is important in the evaluation of the cross sections
because of the $SU(2)$ quantum numbers of the final states, we
expect these results to be sensitive to the $\tgb$ scheme, since
$\tgb$ enters the mixing of the bino and the wino.

\begin{table}[h!]
\begin{center}
\begin{tabular}{lcccccc}
\hline
& & Tree & & $\mathrm{A}_{\tau \tau}$ & $\drbar$ & MH \\
\hline
\multicolumn{1}{l}{$\neuto\neuto\rightarrow W^{+} W^{-}$ [$44\%$]}
                                  & a    &  $+0.81$    &     & $+7.6\%$  &$+12.16\%$& $+29.6\%$ \\
                                  & b    &  $+1.219$   &     & $+0.78\%$ & $+7.1\%$ & $+24.2\%$ \\
\hline
\multicolumn{1}{l}{$\neuto\chargop \rightarrow u \bar{d}$ [$8\%$]}
                                  & a     &  $+15.61$  &      & $+7.2\%$ & $+9.8\%$ & $+18.8\%$ \\
                                  & b     &  $-5.81$   &      & $+5.7\%$ & $+8.3\%$ & $+17.4\%$ \\
\hline
\multicolumn{1}{l}{$\neuto\chargop \rightarrow Z^{0} W^{+}$ [$5\%$]}
                                  & a     &  $+8.26$   &      & $+2.9\%$ & $+4.4\%$ & $+9.7\%$ \\
                                  & b     &  $+1.42$   &      & $-7.3\%$ & $-3.3\%$ & $+10.7\%$ \\
\hline
\multicolumn{1}{l}{$\neuto\neutt \rightarrow W^{+} W^{-}$ [$5\%$]}
                                  & a     &  $+17.81$  &      & $+9.0\%$ & $+11.1\%$ & $+18.2\%$ \\
                                  & b     &  $+11.86$  &      & $+4.8\%$ & $+7.3\%$  & $+16.1\%$ \\
\hline
 $\Omegah$                        &       &    $0.108$   &      & $0.105$  & $0.102$ & $0.097$      \\
 $\frac{\delta \Omegah}{\Omegah}$ &       &            &      & $-2.8\%$ &$-5.6\%$   & $-10.2\%$     \\
\hline
\end{tabular}
\caption{ {\em Mixed-bino scenario: Tree-level values of the
$s$-wave ($a$) and $p$-wave ($b$) coefficients in units $ 10^{-26}
{\rm cm}^{3} {\rm s}^{-1}$, as well as the relative one-loop
corrections in the $\mathrm{A}_{\tau \tau}$, $\drbar$, and MH
scheme. The percentages in the first column refer to the
percentage weight, at tree-level, of that particular channel to
the relic density.}}\label{tab:mixedbino}
\end{center}
\end{table}

\noi Table~\ref{tab:mixedbino} gives in particular the $\tgb$ scheme
dependence. As expected the dependence is not negligible in
particular for the annihilation channel with both LSP in the
initial state. The dependence weakens for the co-annihilation
channels where only one LSP takes part. The $M_H$ scheme is once
again a bad choice showing once
again\cite{baro07,baro09,Sloops-higgspaper} very large
corrections. The difference between the $\mathrm{A}_{\tau \tau}$
and $\drbar$ is about $2\%$ for the co-annihilation channels and
$4\%$ for the annihilation channels. At the end taking into
account the one-loop corrections only to those processes we
studied, which represent $70\%$ off all processes, the correction
on the relic density is about $-3\%$ in the $A_{\tau \tau}$ scheme
and with $\alpha$ defined in the Thomson limit.

\section{A light Higssino scenario}
A pure Higgsino state could give an interesting relic density
and, as the $\neuto\neutt Z^0$ and $\neuto\chargopm W^{\pm}$ are
large, annihilates mainly into WW and ZZ final states. Besides,
as there are three Higgsino like states (two neutralinos and one
chargino) whose mass splitting is small especially if gaugino masses are
large , $\mneuto \simeq \mneuto
\simeq \mchargopm \simeq |\mu|$, co-annihilation between the LSP and the other
higgsino states is important. With such efficient annihilations
the relic density would be small if the Higgsino like
LSP is too light. Nonetheless it gives favourable prospects for dark matter
direct detection. The scenario we have chosen is described in terms of the underlying parameters given in Table~\ref{tab:SUSY2}.
\begin{table}[h]
\begin{center}
\begin{tabular}{ccccccccc}
 \hline
Parameter & $M_1$ & $M_2$ & $\mu$ & $\tb$ & $M_3$ & $M_{\tilde{L},\tilde{Q}}$ & $A_i$ & $M_{A^0}$ \\
 \hline
Value     & 400   & 350   & -250  & 4     & 1000  & 650                       & 0     & 800 \\
\hline
 \end{tabular}\caption{\em Higgsino scenario: Parameters defining our higgsino model with little mixing.
 Masses are in GeV.}\label{tab:SUSY2}
\end{center}
\end{table}

\noi The LSP neutralino with mass $\mneuto = 234$ GeV has a composition
 $\tilde{\chi}_{1}^{0}$ = $0.11\tilde{B} - 0.31\tilde{W} - 0.70\tilde{H}_{1}^{0}
 - 0.63\tilde{H}_{2}^{0}$, indicating it is dominantly a higgsino state. Co-annihilation
between the $\neuto$ and $\chargopm$ occurs since $\mchargopm = 242.9$ GeV. All other particles, including $\neutt$, are heavy
enough and therefore do not take part in the co-annihilation. As
the higgsino component of the neutralino is important, it will
couple mostly to the $W$ and, compared to the mixed-bino case, to
the $Z^0$ boson also through the $\neuto \neutt Z^0$ coupling.

\begin{figure*}[htbp]
\begin{center}
\includegraphics[width=\textwidth]{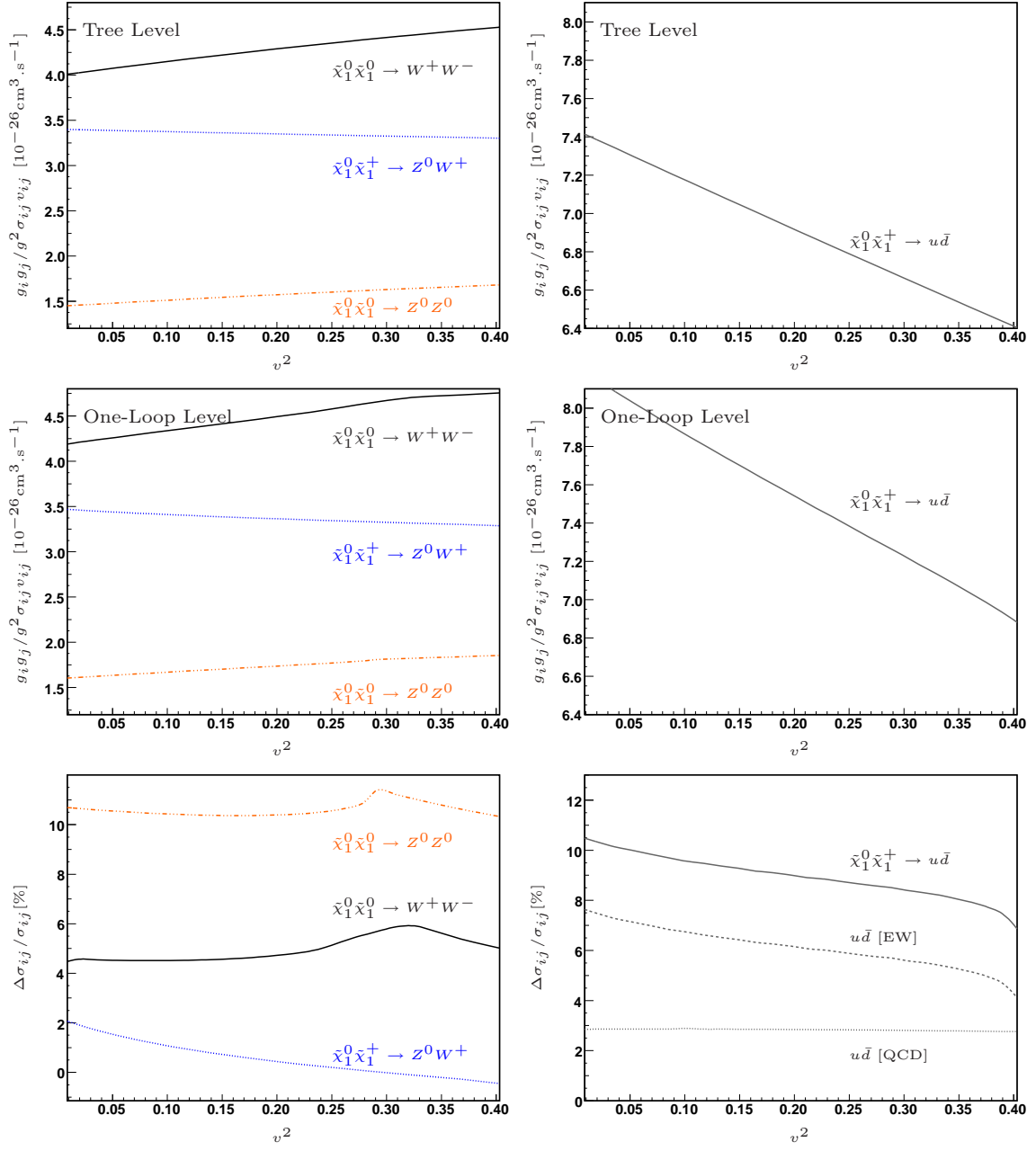}
\caption{\label{figshiggsinogauge} \em Higgsino scenario: The
left/right panel shows the main gauge boson/quark production
cross sections respectively. All the cross sections are normalised
with the corresponding effective degrees of freedom given by
Eq.~(\ref{normgef}) with $x_{F}=27.6$.}
\end{center}
\end{figure*}

\noi Dominant tree-level processes relevant for the computation of the
relic density are the same as in the previous mixed-bino case
except for the co-annihilation between the first two neutralinos
which is Boltzmann suppressed due to their larger mass splitting
and smaller couplings. The dominant processes are
$\neuto\neuto\rightarrow W^{+} W^{-}$ contributing (at tree-level)
$26\%$ to the relic density, $\neuto\chargop \rightarrow u \bar{d}
(c \bar s)$ with $12\%$($12\%$), $\neuto\neuto \rightarrow Z^{0}
Z^{0}$ with
$9\%$ and $\neuto\chargop \rightarrow Z^{0} W^{+}$ with $6\%$. \\
\noindent Our results for the cross sections both at tree-level
and at one-loop are displayed in Fig.~\ref{figshiggsinogauge}.
They are shown for $A_{\tau \tau}$ scheme of $\tgb$ and where the
input for $\alpha$ is in the Thomson limit.

\noi Compared to the mixed bino case, the QCD and EW corrections to
co-annihilation into light quarks are smaller and no cancelation
between the two occurs. The overall correction is almost velocity
independent and ranges between $10\%$ to $8\%$. The corrections to
gauge boson production are smaller for $\neuto\neuto\rightarrow
W^{+} W^{-}$ and $\neuto\chargop \rightarrow Z^0 W^+$ and about
$10\%$ for $\neuto \neuto \rightarrow Z^0 Z^0$.
Fig.~\ref{figshiggsinogauge} shows a very interesting dynamical
effect in the one-loop correction to $\neuto\neuto\rightarrow
W^{+} W^{-}, Z^0 Z^0$ for $v^2 \sim 0.3$. The bumps are in fact
due to the opening of the $\chargop$ threshold in the loop, as can
be checked explicitly for this value of the velocity and the mass
of the LSP compared to that of the $\chargop$. This dynamical
structure can not be described by a simple $a+b v^2$
parametrisation of the cross section. Compared to the bino case we
have studied in the previous section the $\tgb$ scheme dependence
is small. The dependence is shown in Table~\ref{tab:higgsino}
where we also give the results in terms of the $s$-wave and
$p$-wave coefficients for a fit in the range $v^2<0.3$ so that we
avoid the dynamical structure we have just pointed at. The
difference between the $A_{\tau \tau}$ scheme and the $\drbar$
scheme is below $1\%$ for all processes we studied, while the $MH$
scheme gives larger corrections but within $2\%$ compared to the
$\drbar$. For quark production the scheme dependence is even
negligible. The overall ${\cal {O}}(\alpha)$ corrections in this
scenario, though not negligible, are not that large with $\alpha$
defined in the on-shell scheme in the Thomson limit. Moreover
corrections coming from boxes and final state radiation are often
dominant. This suggests that to grab most of the radiative
corrections a full calculation is needed. Within our approach of
not correcting the processes that contribute less than $5\%$ to
the relic density, the processes we have considered contribute in
total only $65\%$. In this approach we find a correction to the
relic density of $-2.5\%$ in the $A_{\tau \tau}$ scheme and
$-2.4\%$ in $\drbar$. In the $MH$ scheme the correction is little
changed to $-3.3\%$.
\begin{table}[hbtp]
\begin{center}
\begin{tabular}{lcccccc}
\hline
& & Tree & & $\mathrm{A}_{\tau \tau}$ & $\drbar$ & MH \\
\hline
\multicolumn{1}{l}{$\neuto\neuto\rightarrow W^{+} W^{-}$ [$26\%$]}
                                  & a    &  $+11.84$  &     & $+4.3\%$   & $+5.1\%$  & $+6.8\%$ \\
                                  & b    &  $+4.17$   &     & $+12.7\%$  & $+13.4\%$ & $+14.9\%$ \\
\hline
\multicolumn{1}{l}{$\neuto\chargop \rightarrow u \bar{d}$ [$12\%$]}
                                  & a     &  $+15.28$ &      & $+6.8\%$  & $+7.0\%$  & $+7.3\%$ \\
                                  & b     &  $-5.31$  &      & $+30.4\%$ & $+30.7\%$ & $+31.3\%$ \\
\hline
\multicolumn{1}{l}{$\neuto\neuto \rightarrow Z^{0} Z^{0}$ [$9\%$]}
                                  & a     &  $+4.28$  &      & $+10.4\%$ & $+9.6\%$  & $+7.8\%$ \\
                                  & b     &  $+1.83$  &      & $+12.7\%$ & $+12.0\%$ & $+10.5\%$ \\
\hline
\multicolumn{1}{l}{$\neuto\chargop \rightarrow Z^{0} W^{+}$ [$6\%$]}
                                  & a     &  $+6.99$  &      & $+1.7\%$  & $+2.1\%$  & $+2.9\%$ \\
                                  & b     &  $-0.51$  &      & $+85.6\%$ & $+86.5\%$ & $+88.4\%$ \\
\hline
 $\Omegah$                        &       &    $0.00931$&      &  $0.00909$  &  $0.00908$  &   $0.00904$       \\
 $\frac{\delta \Omegah}{\Omegah}$ &       &             &      &  $-2.4\%$ &  $-2.5\%$ &   $-3.3\%$  \\
\hline
\end{tabular}
\caption{ {\em Higgsino scenario: Tree-level values of the
$s$-wave ($a$) and $p$-wave ($b$) coefficients in units $ 10^{-26}
{\rm cm}^{3} {\rm s}^{-1}$ in the higgsino scenario, as well as
the relative one-loop corrections in the $\atts$, $\drbar$, $\mhs$
scheme. The percentages in the first column next to the process
refer to the percentage weight, at tree-level, of that particular
channel to the relic density. The fit into $a$ and $b$ is done in
the range $0<v^2<0.3$.}}\label{tab:higgsino}
\end{center}
\end{table}

\section{A light wino scenario}
Models with a light wino as the dark matter candidate occur in
theories like AMSB\cite{AMSB}, string
compactifications\cite{Acharya:2008zi} and also
split-supersymmetry\cite{WellsPeV,Split}. The advantage of a light
wino is that it has a large annihilation cross section, relevant
for indirect detection, but the main drawback is that it predicts
a small thermal relic abundance in the standard cosmological
scenario and non-thermal production has to be invoked to recover
the correct relic density.
The underlying parameters of the model are given in
Table~\ref{tab:SUSY4}.

\begin{table}[bth]
\begin{center}
\begin{tabular}{ccccccccccc}
 \hline
Parameter & $M_1$ & $M_2$ & $\mu$ & $\tb$ & $M_3$ & $M_{\tilde{u}_L}$ & $M_{\tilde{e}_L}$ & $M_{\tilde{u}_R,\tilde{e}_R}$ & $A_i$ & $M_{A^0}$ \\
 \hline
Value     & 550   & 210   & -600  & 30    & 1200  & 387               & 360                &
800                           & 0     & 700  \\
\hline
 \end{tabular}\caption{\em Light-wino scenario: Values of the fourth SUSY set of input
 parameters. Masses are in GeV.}\label{tab:SUSY4}
\end{center}
\end{table}

\noi The LSP is now essentially wino with a composition
$\tilde{\chi}_{1}^{0}$ = $0.005\tilde{B} - 0.99\tilde{W} -
0.15\tilde{H}_{1}^{0} - 0.05\tilde{H}_{2}^{0}$ and mass $\mneuto = 206.6$ GeV.
 The LSP is highly degenerate with the $\chargopm$,
their mass difference is $\Delta m \simeq 0.05$ GeV and
consequently sizeable co-annihilations occur in the determination
of the relic density. With so small mass difference,
co-annihilation processes are important. Products of
annihilation/co-annihilation processes are into gauge bosons (and
some light quarks). The dominant processes are the following:
$\neuto\neuto\rightarrow W^{+} W^{-}$ [$13\%$],
$\chargop\chargop\rightarrow W^{+} W^{+}$ [$12\%$],
$\neuto\chargop\rightarrow Z^{0} W^{+}$ [$12\%$],
$\chargop\chargom \rightarrow Z^{0} Z^{0}$ [$7\%$],
$\chargop\chargom\rightarrow W^{+} W^{-}$ [$7\%$], $\neuto\chargop
\rightarrow u \bar{d}$ [$7\%$]. Note in passing that we
have taken a large value of $\tgb$. \\

\noi Before we present our results, a word about the renormalisation
scheme and the choice of input parameters especially as concerns
the neutralino/chargino sector is in order. The results we will
show are based on taking the mass of the LSP as input (beside the
masses of the charginos). One might argue that this is not optimal
in order to reconstruct the system $M_1,M_2,\mu$, especially for
extracting $M_1$ which is sensitive to the bino-component. One
might even expect that at one-loop this scheme would not be suitable since the $N_{11}$
element of the orthogonal matrix in the neutralino sector is very
small leading to a large contribution from the counterterm
$\delta M_1$, see Eq.~(\ref{dm1dm2dmu}). For this reason we have
been careful in also taking the scheme where the input parameters
are $(m_{\tilde{\chi}_{2}^{0}}, m_{\tilde{\chi}_{1}^{+}},
m_{\tilde{\chi}_{2}^{+}})$.
\begin{figure*}[h]
\begin{center}
\includegraphics[width=0.45\textwidth]{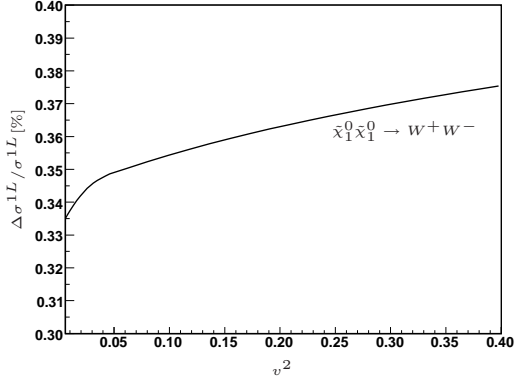}
\caption{\label{figsdifflight} \em Relative difference for
$\tilde{\chi}_{1}^{0}\tilde{\chi}_{1}^{0}\rightarrow W^{+}W^{-}$
between the scheme with $(m_{\tilde{\chi}_{1}^{0}},
m_{\tilde{\chi}_{1}^{+}}, m_{\tilde{\chi}_{2}^{+}})$ and
$(m_{\tilde{\chi}_{2}^{0}}, m_{\tilde{\chi}_{1}^{+}},
m_{\tilde{\chi}_{2}^{+}})$ as function of relative velocity. $\tb$
is within the $A_{\tau \tau}$ scheme.}
\end{center}
\end{figure*}
Fortunately, as we can see in Fig. \ref{figsdifflight} for the
process $\tilde{\chi}_{1}^{0}\tilde{\chi}_{1}^{0} \rightarrow
W^{+}W^{-}$, the difference between the two schemes is within less
than $0.4\%$. We have checked that for other processes in this
scenario the difference is also negligible.

\begin{figure*}[htbp]
\begin{center}
\includegraphics[width=\textwidth]{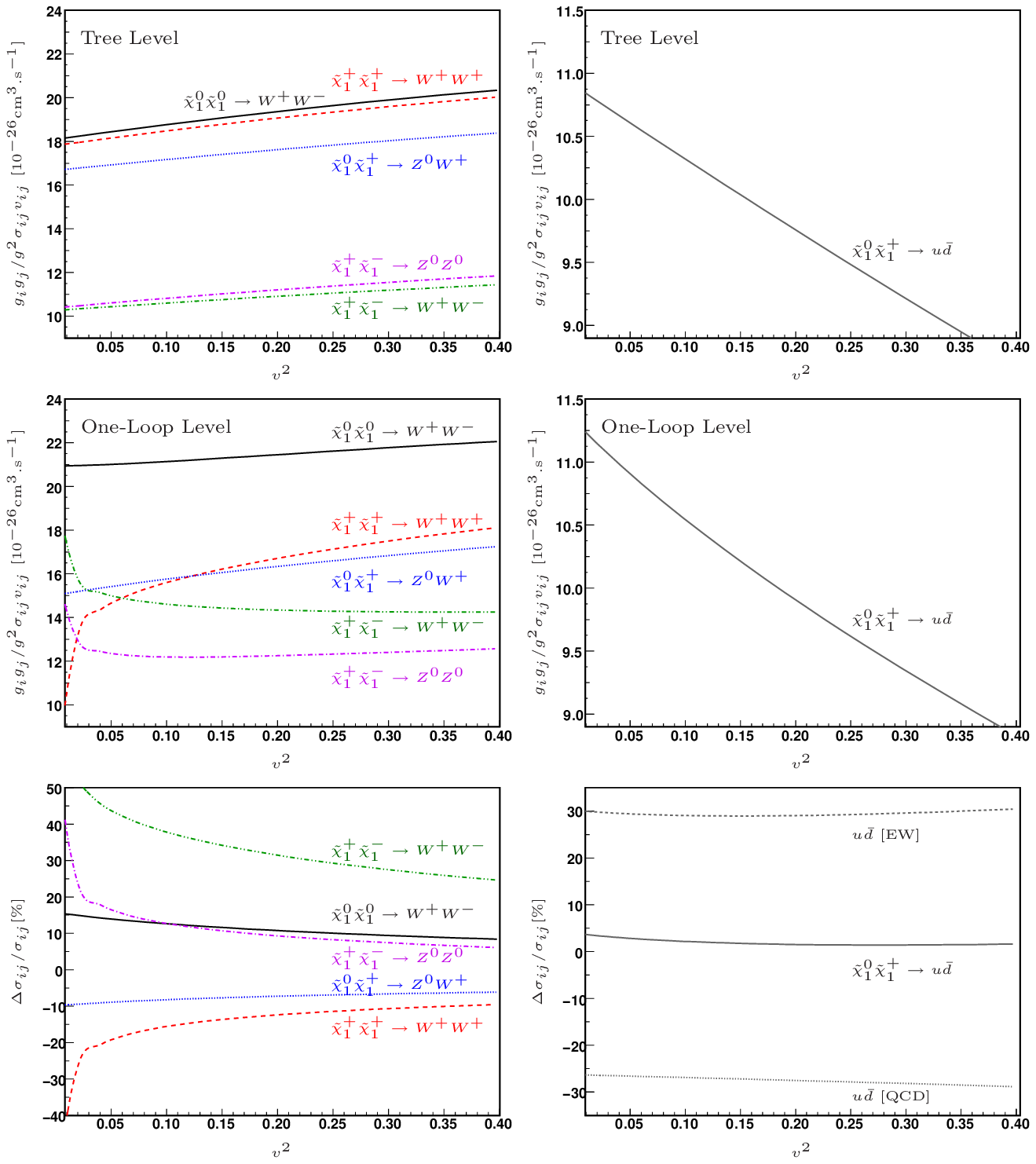}
\caption{\label{figslightwinogauge} \em Light-wino scenario: The
left/right panel shows the main gauge boson/quark production
cross sections respectively. All the cross sections are normalised
with the corresponding effective degrees of freedom given by
Eq.~(\ref{normgef}) with $x_{F}=29.9$.}
\end{center}
\end{figure*}

\begin{figure*}[h]
\begin{center}
\includegraphics[width=0.45\textwidth]{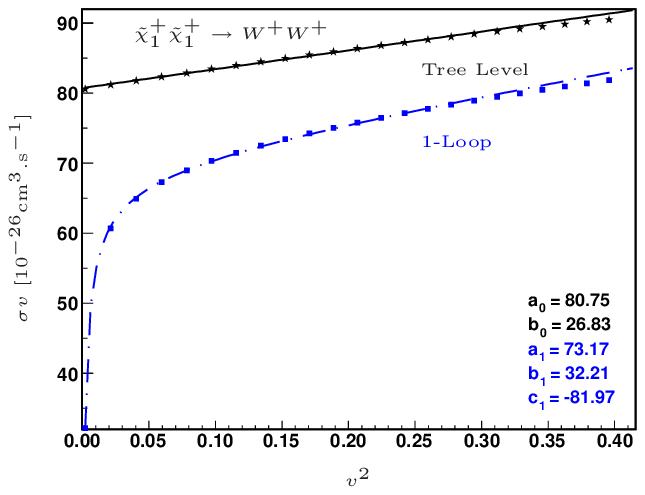}
\caption{\label{fitlightwino} \em Lightwino scenario. Fits to the
$s$-wave, $p$-wave and Sommerfeld factors for
$\chargop\chargop\rightarrow W^{+} W^{+}$. In particular note that
the fit in the parametrisation of Eq.~(\ref{fit-1l-sommerfeld})
gives $c_1/a_0=1.015$. }
\end{center}
\end{figure*}

\begin{figure*}[h!]
 \begin{center}
\includegraphics[width=0.49\textwidth]{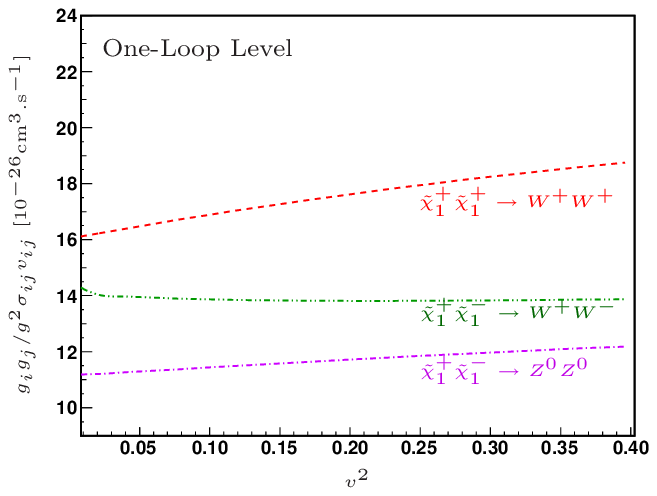}
\caption{\label{figslightwinogaugenoQED} \em {Light-wino scenario:
Results for one-loop corrections in the $\Att$ scheme where the
QED Sommerfeld effect has been subtracted.}}
 \end{center}
\end{figure*}

\noi The tree-level cross sections and the full one-loop corrections
are shown in Fig.~\ref{figslightwinogauge} as a function of the
relative velocity. We note that, at tree-level, these cross
sections are $s$-wave dominated. For bosonic final states the
velocity dependence is modest, compared to the co-annihilation
into quarks. For the latter, the electroweak and QCD corrections
to $\neuto\chargop \rightarrow u \bar{d}$ are relatively large, of
order $30\%$, but they almost cancel each other. The overall
correction is about $+5\%$ and practically independent of the
velocity. The annihilation process and the $\neuto\chargop$
co-annihilation processes show an almost constant correction of
order $10\%$. The co-annihilation processes show an interesting
behaviour in the case where both co-annihilating particles are
charged, the cross sections reveal a very large correction at very
small relative velocity. This correction is the one-loop
manifestation of the non-relativistic Coulomb-Sommerfeld
effect\cite{textbook-Sommerfeld}. With the tree-level cross
section denoted as $\sigma_0$ and $\sigma_0 v = a_0 + b_0 v^2$, at
vanishing relative velocity the one loop cross section for
chargino annihilation, $\sigma_{\mathrm{Coul}}^{\mathrm{1-loop}}$
is such that
\begin{eqnarray*}
 \frac{\sigma_{\mathrm{Coul}, v \rightarrow 0}^{\mathrm{1-loop}}}{\sigma_0} & = &\begin{cases}
                                      +{\frac{\pi\alpha}{v}} \quad \mathrm{for} \quad \chargopm\chargomp \\
                              -{\frac{\pi\alpha}{v}} \quad \mathrm{for} \quad \chargopm\chargopm
                                      \end{cases}
\end{eqnarray*}
We thus expect the one-loop cross section $\sigma_1$ to be
\begin{equation}
\label{fit-1l-sommerfeld}
 \sigma_1 v = a_1 + b_1 v^2 + \pi\alpha c_1/v \quad \mathrm{with} \quad c_1 = \pm
 a_0.
\end{equation}
To exactly quantify the Sommerfeld effect in our calculation, we
have also fitted the one-loop cross section in the form of
Eq.~(\ref{fit-1l-sommerfeld}). An example of such an exercise is
given in Fig.~\ref{fitlightwino} for $\chargop\chargop\rightarrow
W^{+} W^{+}$.
We see therefore that our calculation captures this effect
extremely well, indeed we obtain here that $c_1/a_0=1.015$ which
indeed very close to the analytical result, $c_1/a_0=1$. This is
important because this effect needs to be summed up to all orders.
In our approach we will therefore subtract it from the one-loop
correction and replace it by the resummed all order result in the
final result. The result of this subtraction on the processes
$\chargop\chargom \rightarrow Z^{0} Z^{0}$, $\chargop\chargop
\rightarrow W^{+} W^{+}$ and $\chargop\chargom \rightarrow W^{+}
W^{-}$ is shown in Fig.~\ref{figslightwinogaugenoQED}

\begin{figure}[htb]
\begin{center}
 \includegraphics[width=0.45\textwidth]{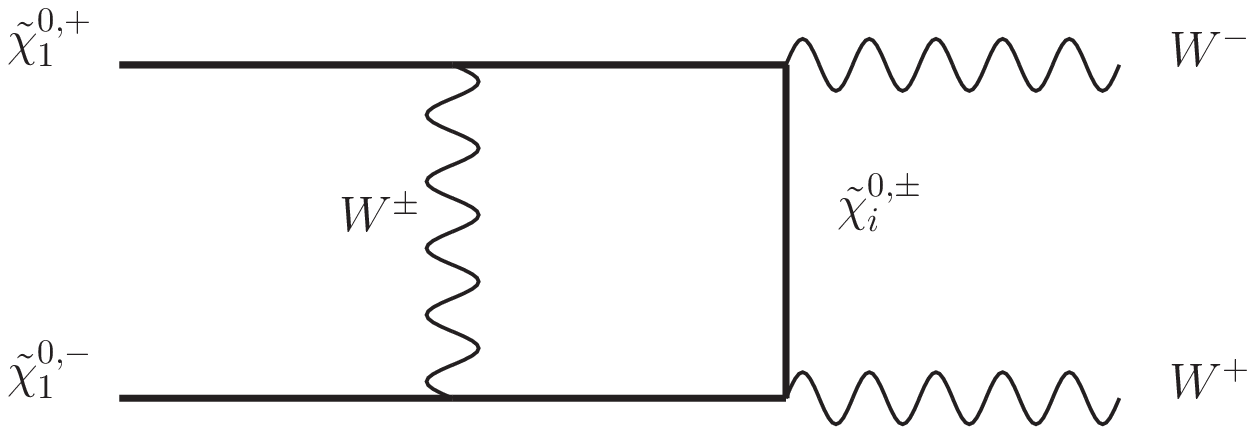}\caption{\label{boxcharg}
{\em Example of a box diagram giving rise to the electroweak
equivalent of a Sommerfeld effect.}}
\end{center}
\end{figure}

\begin{figure}[hbt]
\begin{center}
\includegraphics[width=0.45\textwidth]{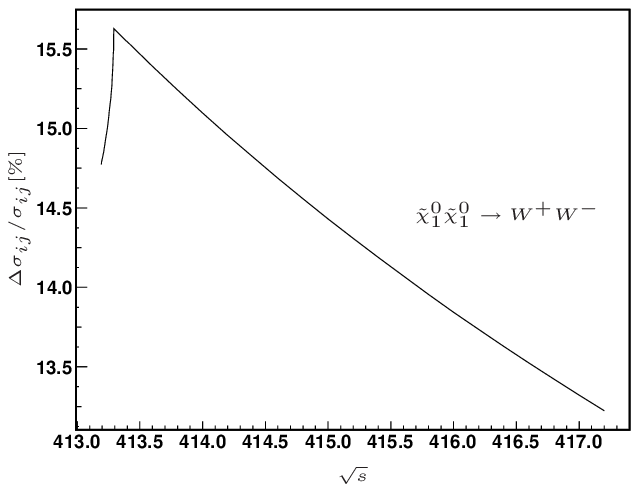}
 \caption{\label{o1o1WWsmallv}{
  \em
 Light-wino scenario: the kink in $\neuto \neuto \rightarrow
  W^{+} W^{-}$ at small relative velocity corresponding to the opening of the
  threshold for $\chargopm$ at
$\sqrt{s} = 2\times\mchargopm = 2\times206.646=413.3 $ GeV}}
\end{center}
\end{figure}

\noi As one can see once the Coulomb-Sommerfeld contribution is
removed, one is left with a smooth cross section which is almost
velocity independent.\\
\noindent Looking carefully at the results for
$\chargop\chargom \rightarrow W^{+} W^{-}$ we note that there is
still a slight increase at small $v$. This is a residual effect of
the weak Sommerfeld contribution, see Fig.~\ref{boxcharg},
mediated by a charged $W$ that it is noticeable even for a not too
heavy chargino. In fact a similar effect is also present in
$\neuto\neuto \rightarrow W^{+}W^{-}$. Zooming in on the region of
small relative velocity we see a kink, see
Fig.~\ref{o1o1WWsmallv}, around $\sqrt{s} \simeq 413.3$ GeV
corresponding to $v \simeq 0.04$ which corresponds to the opening
of the threshold of chargino production. In $\chargop\chargom
\rightarrow W^{+} W^{-}$ we only see the tail of the opening of
the threshold.

\noi Apart from these interesting but numerically small features, let
us mention that the $\tb$ scheme dependence is negligible, it is
below $0.1\%$. Our results show that corrections to the
individual cross sections can get large at all relative
velocities, even after subtracting the QED Sommerfeld effect. For
example, $\chargop\chargom\rightarrow W^{+} W^{-}$ gets about
$+30\%$ correction, while both $\chargop\chargop\rightarrow W^{+}
W^{+}$ and $\neuto\chargop\rightarrow Z^{0} W^{+}$ get a $-10\%$
correction. The dominant cross section $\neuto\neuto\rightarrow
W^{+} W^{-}$ receives a correction of about $+15\%$. The
corrections for the other processes are more modest. These
corrections are within the on-shell scheme by using $\alpha(0)$.
We see that had we used $\alpha(M_Z^2)$ the correction to
$\neuto\neuto\rightarrow W^{+} W^{-}$ would be small, but this is
not true for most of the other cross sections where genuine
corrections, including hard radiation effects are important and
must be taken into account. This said, when we combine all the
cross sections, taking into account their statistical weight,
substantial cancelations occur between the different
contributions.
\begin{table}[htb]
\begin{center}
\begin{tabular}{lcccc}
\hline
& & Tree & & Full ${\cal {O}}(\alpha)$ \\
\hline \multicolumn{1}{l}{$\neuto\neuto\rightarrow W^{+} W^{-}$
[$13\%$]}
                                  & $a$    &  $+161.8$ &     & $+14.8\%$  \\
                                  & $b$    &  $+53.52$ &     & $-48.7\%$  \\
\hline \multicolumn{1}{l}{$\chargop\chargop\rightarrow W^{+}
W^{+}$ [$12\%$]}
                                  & $a$    &  $+80.75$  &     & $-9.4\%$     \\
                                  & $b$    &  $+26.83$  &     & $+20.1\%$    \\
                                  & $c$    &            &     & $-81.97$     \\
\hline \multicolumn{1}{l}{$\neuto\chargop\rightarrow Z^{0} W^{+}$
[$12\%$]}
                                  & $a$    &  $+37.50$  &     & $-9.5\%$ \\
                                  & $b$    &  $+10.15$  &     & $+31.6\%$ \\
\hline \multicolumn{1}{l}{$\neuto\chargop \rightarrow u \bar{d}$
[$7\%$]}
                                  & $a$    &  $+24.44$  &     & $+3.17\%$ \\
                                  & $b$    &  $-12.62$  &     & $+16.3\%$ \\
\hline \multicolumn{1}{l}{$\chargop\chargom \rightarrow Z^{0}
Z^{0}$ [$7\%$]}
                                  & $a$    &  $+47.08$  &     & $+7.1\%$    \\
                                  & $b$    &  $+17.71$  &     & $-29.0\%$   \\
                                  & $c$    &            &     & $+47.1$     \\
\hline \multicolumn{1}{l}{$\chargop\chargom\rightarrow W^{+}
W^{-}$ [$7\%$]}
                                  & $a$    &  $+46.49$  &     & $+34.0\%$    \\
                                  & $b$    &  $+14.01$  &     & $-104.4\%$ \\
                                  & $c$    &            &     & $+53.34$     \\
\hline
 $\Omegah$                        &        & $0.00215$  &     & $0.00211$  \\
 $\frac{\delta \Omegah}{\Omegah}$1-loop &        &            &     & $-1.9\%$   \\
$\frac{\delta \Omegah}{\Omegah}$ with resum. Sommerfeld & & & &
$-1.9\%$ \\ \hline
\end{tabular}
\caption{ {\em Light-wino scenario. The table summarises the
effect of the full order corrections on the dominant processes
that contribute more than $5\%$ to the relic density. The relative
contribution is given in [ ] next to the process. The tree-level
cross sections are given through the fit $\sigma v = a + bv^2$ in
the range $0 < v^2 <0.3$. At the one-loop level, The fitting
function is then $\sigma v = a + bv^2 + c\pi\alpha/v$. The
coefficients $a$, $b$ and $c$ are given units of $10^{-26}{\rm
cm}^3 {\rm s}^{-1}$. The relic density is calculated by taking
into account all other processes, which however are not corrected
at one-loop. The Table also gives the correction after summing the
$1/v$ QED contribution at all orders. As the $t_{\beta}$-scheme
dependence is less than $0.1\%$, only one $t_{\beta}$ scheme
$A_{\tau \tau}$ is presented. }}\label{tab:light-wino}
\end{center}
\end{table}
Add to this that the cross sections we have considered account for
about only $65\%$ of all cross sections contributing to the relic
density, since we have not considered those contributing
individually less than $5\%$, we find a quite modest (within the
$\alpha(0)$ scheme) correction to the relic density of about
$-2\%$, see ~Table~\ref{tab:light-wino}. This full one-loop
correction is practically unchanged if we instead resum the $1/v$
Sommerfeld effect. A similar result was found when we studied
$\tilde{\tau}$ co-annihilation\cite{baro07}. This is due to the
fact that temperature effects provide a cut-off and the $1/v$
enhancement is tamed after thermal averaging, $\propto
\int_0^{\infty} ({\rm d}v \; v^2 \; e^{-v^2/4 x} ) \; (\sigma v)$.
Our results are summarised in Table~\ref{tab:light-wino}. The
results are presented in terms of the $s$-wave and $p$-wave
coefficients as well as the Sommerfeld $1/v$ coefficient. The
correction to the relic density is given for the full one-loop,
including the one-loop $1/v$ threshold correction, as well as
after resumming the $1/v$ contributions. Another word of warning
about the interpretation of the corrections in terms of the
$s$-wave and $p$-wave coefficients ($a$ and $b$). The corrections
to the $p$-wave coefficients may seem very large here, especially
if the corresponding correction to $a$ is large. This is not an
indication that the radiative correction on the total cross is
very large. Indeed all the cross sections here are $s$-wave
dominated, so that the correction on the $s$-wave is a good
measure of the total correction and when modulated with the
statistical weight gives a good approximation to the correction on
the relic density.

\section{Conclusions}
Extraction of the relic density will soon provide a measurement of
this quantity at the $1\%$ level. On the theoretical side one must
therefore provide a prediction which is at least as precise. In
particular, if the particle physics component in terms of
computation of the annihilations and co-annihilations cross
sections are under control, one can indirectly test the cosmology
of the Universe. With this precise measurement we can even gain
insight into the particle physics model that could be combined
with measurements at the colliders. The work in this paper
continues the investigations we have made\cite{baro07} concerning
the impact of the radiative corrections on the annihilation and
co-annihilation cross sections of a neutralino dark matter in the
MSSM. Here the emphasis is on processes with dominant gauge boson
production channels. We have considered three models with
relatively light LSP in the $100$-$200$ GeV range, {\it i}) a
dominantly bino with some admixture of wino, {\it ii}) a higgsino
like and a {\it iii}) wino like LSP. Our study shows that it is not
easy to find a general common feature of the radiative
corrections. For example, within the same $\tgb$ scheme and for
relative velocities relevant for the evaluation of the relic
density, the dominant process $\neuto\neuto\rightarrow W^{+}
W^{-}$ gets about $15\%$ correction in the wino case but only
$4\%$ in the higgsino case, while the bino is in between. Also the
corrections we have just quoted are within a scheme where the
electromagnetic constant is defined at low energy in the Thomson
limit. A naive use of $\alpha(M_Z^2)$ would suggest that most of
the corrections in the dominant process in the wino case is
absorbed, but this would not be true for the other processes nor
for the the other scenarios. This still does not take into account
the effects of final state radiation. For example in the same wino
scenario, the ${\cal {O}}(\alpha)$ correction to
$\chargop\chargom\ \rightarrow W^{+} W^{-}$ is large and reaches
about $30\%$. In general the corrections to the different
contributing processes for the same scenario can be quite
different, in the case of the wino the overall effect on the relic
is a cancelation of the corrections between the different
processes. With this in mind and the fact that we did not correct
processes that, individually, contribute less than $5\%$ to the
relic, we find that the overall effect on the relic is small,
$-2\%$ in the on-shell scheme with $\alpha$ in the Thomson limit.
In this paper we have also pursued our investigation of the effect
of the $\tgb$ scheme dependence on many observables, not
necessarily dark matter annihilation. We confirm once more that
the $MH$ scheme is not appropriate while $\drbar$ and $A_{\tau
\tau}$ give generally similar results. In the scenarios we have
studied, in fact the scheme dependence is an issue only for the
bino case. This could have been expected as the bino couples to
$W$'s only through mixing where $\tgb$ is important. We have also
uncovered in the case of co-annihilation electromagnetic
Sommerfeld effects for vanishingly small relative velocity.
However the result of the full one-loop and that of resumming this
effect is not noticeable at the level of the relic density
evaluation, thermal averaging smoothes out the effect. Although
the effect is numerically quite small, in the case of the wino we
noticed the effect of the electroweak Sommerfeld enhancement. This
will become more important for higher wino masses that we will
study in another paper. To sum up, it is important to stress that
we now have the tools to perform automated calculations relevant
for a precise evaluation of the dark matter annihilation cross
sections.

\vspace{1cm}
\noi {\bf \large Acknowledgments} \\
We would like to thank Sasha Pukhov for helping us with the
interface of {\tt SloopS} with {\tt micrOMEGAs}. We benefited from
many discussions with him and Genevieve Belanger. This work is
part of the French ANR project, ToolsDMColl BLAN07-2 194882 and is
supported in part by the GDRI-ACPP of the CNRS (France). This work
is also supported in part by the European Community's Marie-Curie
Research Training Network under contract MRTN-CT-2006-035505
``Tools and Precision Calculations for Physics Discoveries at
Colliders'', the DFG SFB/TR9 ``Computational Particle Physics'', and
the Helmholtz Alliance ``Physics at the Terascale''.



\end{document}